\documentclass[aps,prl,twocolumn,showpacs,showkeys,groupedaddress]{revtex4}

\usepackage{graphicx}

\begin{document}


\title{Model-independent assessment of current direct searches for spin-dependent dark matter}


\author{F. Giuliani}
\affiliation{Centro de Fisica Nuclear, Universidade de Lisboa, 1649-003 Lisboa, Portugal}


\date{March 30, 2004}

\begin{abstract}
I evaluate the current results of spin-dependent weakly interacting massive particle (WIMP) searches within a model-independent framework, showing the most restrictive limits to date derive from the combination of xenon and sodium iodide experiments. The extension of this analysis to the case of positive signal experiments is elaborated.
\end{abstract}


\maketitle



%

It is well known that the spin-dependent WIMP-nucleus interaction is quenched by spin pairing effects in the nucleus. For this reason, ideal nuclei for a spin-dependent WIMP search would be both odd Z and odd neutron number (N), hence even mass number A. Unfortunately, actual detectors use only odd A nuclei, so that their response is dominated by the WIMP-proton (WIMP-neutron) interaction only. Interpretation of the results has customarily proceeded by assuming the WIMP interacts with only the dominant group of nucleons. In this "odd group" approximation, an experiment using only odd Z isotopes cannot constrain a theory predicting a WIMP which couples with neutrons only, i.e., the exclusion results quoted by such experiment are WIMP model dependent. This is particularly true with the current WIMP candidate: the neutralino is a superposition of higgsinos and gauginos, whose weightings determine the coupling strengths.

On the other hand, some odd Z nuclei, like F and I, exhibit a non-negligible neutron group spin, making it reasonable to go beyond the "odd group" approximation. Recently, some experiments began to adopt this approach \cite{naiad,danai,Tokyo,plb}, using (except for DAMA) a full spin-dependent treatment proposed by Tovey et. al.\cite{Tovey}.

Here, the impact of this framework on the interpretation of current experiments is considered, first in the case of null results and then for the case of a positive WIMP signal detection such as DAMA/NaI. This letter focuses on experiments that have already published spin-dependent limits; others will be considered in a forthcoming paper. The combination of xenon- and sodium iodide-based experiments is seen to provide the most restrictive current limits on the existence of spin-dependent WIMP dark matter, with the fluorine-based (F-based) experiments offering a good possibility for near-term improvements \cite{privcom}.

At tree level, the general (zero momentum transfer) WIMP-nuclide spin-dependent elastic scattering cross section $\sigma_{A}$ for a nucleus of mass number A is \cite{Lewin,Engel,Kurylov}:

\begin{equation}
\sigma_{A}=\frac{32}{\pi}G_{F}^{2}\mu_{A}^{2}(a_{p}<S_{p}>+a_{n}<S_{n}>)^{2}\frac{J+1}{J}.
\label{basic}
\end{equation}

\noindent where $<S_{p,n}>$ are the expectation values of the proton (neutron) group's spin, $G_{F}$ is the Fermi coupling constant, $a_{p,n}$ are the effective proton (neutron) coupling strengths which appear in the effective lagrangian, $\mu_{A}$ is the WIMP-nuclide reduced mass, and J is the total nuclear spin. A discussion on the generality of Eqn. (\ref{basic}) can be found in Ref. \cite{Kurylov}.
When limits on $\sigma_{A}$ are measured, experiments usually quote limits only for one nucleon. This is done by calculating, for each sensitive nuclide, one of the following cross sections:

\begin{equation}
\sigma_{p,n}^{lim(A)}=\frac{3}{4}\frac{J}{J+1}\frac{\mu_{p,n}^{2}}{\mu_{A}^{2}}\frac{\sigma_{A}^{lim}}{<S_{p,n}>^{2}}
\label{sigpnlim}
\end{equation}

\noindent where $\sigma_{A}^{lim}$  is the upper limit on $\sigma_{A}$ obtained from experimental data, $\mu_{p,n}$ is the WIMP-proton (WIMP-neutron) reduced mass, and $\sigma_{p,n}^{lim(A)}$ are the proton and neutron cross section limits when $a_{n,p} = 0$ respectively. The $\sigma_{A}^{lim}$ is evaluated by attributing the entire counting rate to the nuclide A, so $\sigma_{p,n}^{lim(A)}$ are overestimated by a factor $f_{A}^{-1}$, where $f_{A}$ is the isotopic fraction ($\frac{\text{no. of nuclei of isotope A}}{\text{total no. of sensitive nuclei}}$) of the isotope. Since $\sum_{A}{f_{A}}=1$, and $f_{A}\propto \frac{1}{\sigma_{p,n}^{lim(A)}}$, the constant of proportionality is taken as a more refined WIMP-proton (WIMP-neutron) cross section limit $\sigma_{p,n}^{lim}$, obtained from Eqn. (\ref{sigpnlim}) with the relation \cite{Lewin} 
$ \frac{1}{\sigma_{p,n}^{lim}}=\sum_{A}{\frac{1}{\sigma_{p,n}^{lim(A)}}}$.

Eqn. (\ref{basic}) for the single proton (neutron) cross section $\sigma_{p}$ ($\sigma_{n}$) reads:

\begin{equation}
\sigma_{p,n}=\frac{32}{\pi}G_{F}^{2}\mu_{p,n}^{2}\frac{3}{4}a_{p,n}^{2}.
\label{sigpn}
\end{equation}

\noindent Thus, given the $\sigma_{p,n}^{lim(A)}$, model-independent limits on spin-dependent WIMP-nucleon interaction can be formulated either in terms of the $\sigma_{p,n}$ or the coupling strengths $a_{p,n}$. Although equivalent formulations, I discuss in $a_{p,n}$, with both elaborated in a forthcoming paper.

Substituting Eqns. (\ref{basic}) - (\ref{sigpn}) into the obvious relation $\frac{\sigma_{A}}{\sigma_{A}^{lim}} \leq 1$ and taking $m_{n} \approx m_{p}$, it can be shown that \cite{Tovey}

\begin{equation}
(\frac{a_{p}}{\sqrt{\sigma_{p}^{lim(A)}}}\pm \frac{a_{n}}{\sqrt{\sigma_{n}^{lim(A)}}})^{2} \leq \frac{\pi}{24G_{F}^{2}\mu_{p}^{2}}
\label{mono}
\end{equation}

\noindent where the sign of the addition in parenthesis is that of the $\frac{<S_{n}>}{<S_{p}>}$ ratio. The allowed values of $a_{p}$ and $a_{n}$ are constrained to the inside of a degenerate conic in the $a_{p}-a_{n}$ plane, namely to a band between two parallel straight lines, whose stiffness is $-\frac{<S_{n}>}{<S_{p}>}$ . This indicates the inability of a single nuclide experiment to fully constrain all tree level theoretical scenarios.

For detectors with more than one active nuclide, as pointed out above, $\sigma_{p,n}^{lim(A)}$ are overestimated by a factor $f_{A}^{-1}$. Consequently the limits of Eqn. (\ref{mono}) become, for each nuclide, 

\begin{equation}
(\frac{a_{p}}{\sqrt{\sigma_{p}^{lim(A)}}}\pm \frac{a_{n}}{\sqrt{\sigma_{n}^{lim(A)}}})^{2} \leq \frac{\pi}{24G_{F}^{2}\mu_{p}^{2}}f_{A}
\label{monofrac}
\end{equation}

Since $\sum_{A}{f_{A}}=1$ , then:

\begin{equation}
\sum_{A}{(\frac{a_{p}}{\sqrt{\sigma_{p}^{lim(A)}}}\pm \frac{a_{n}}{\sqrt{\sigma_{n}^{lim(A)}}})^{2}} \leq \frac{\pi}{24G_{F}^{2}\mu_{p}^{2}}
\label{multi}
\end{equation}

\noindent where the sum is intended over the nuclear species present in the detector's sensitive volume.

Eqn. (\ref{multi}) removes the conic degeneration of Eqn. (\ref{mono}), even if a nuclide that has little sensitivity to the WIMP-neutron interaction, \textit{e.g.} Na or Cl, is included. The allowed region is now a conic, and the exclusion results of multinuclide detectors are usually ellipses in the $a_{p}-a_{n}$ plane.

Since $\sigma_{p,n}^{lim(A)}$ are functions of the WIMP mass $M_{\chi}$ only, Eqn. (\ref{multi}) is a three parameter relation, which yields a 3D allowed region, shaped as a tube with varying elliptical cross section transverse to the mass axis. Results can be reported \cite{Tovey,naiad,Tokyo,plb} by plotting $\sigma_{p,n}^{lim(A)}$ vs $M_{\chi}$, plus allowed regions for selected $M_{\chi}$ values. This is shown for

 \begin{figure*}
 \includegraphics[width=7.5 cm]{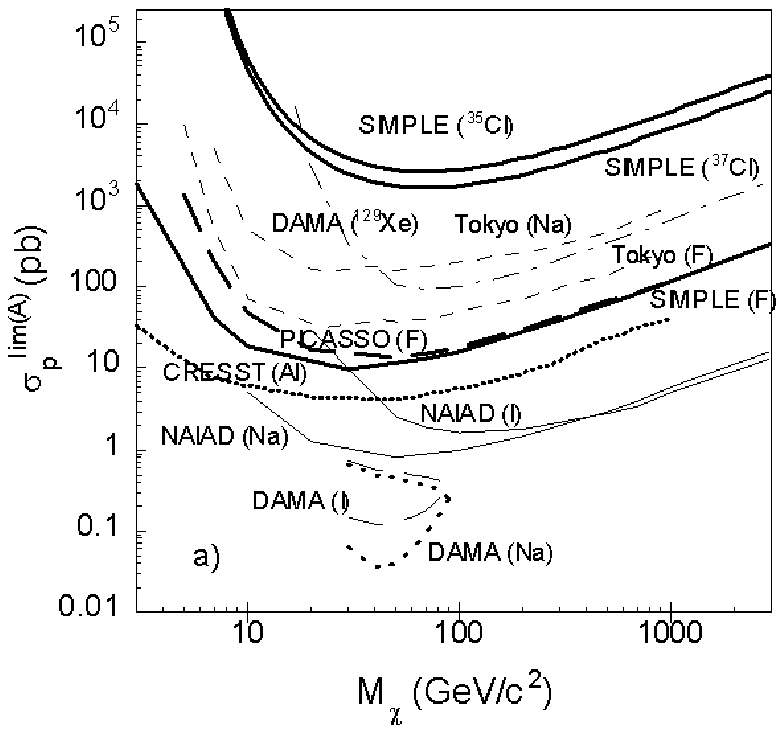}
 \includegraphics[width=7.5 cm]{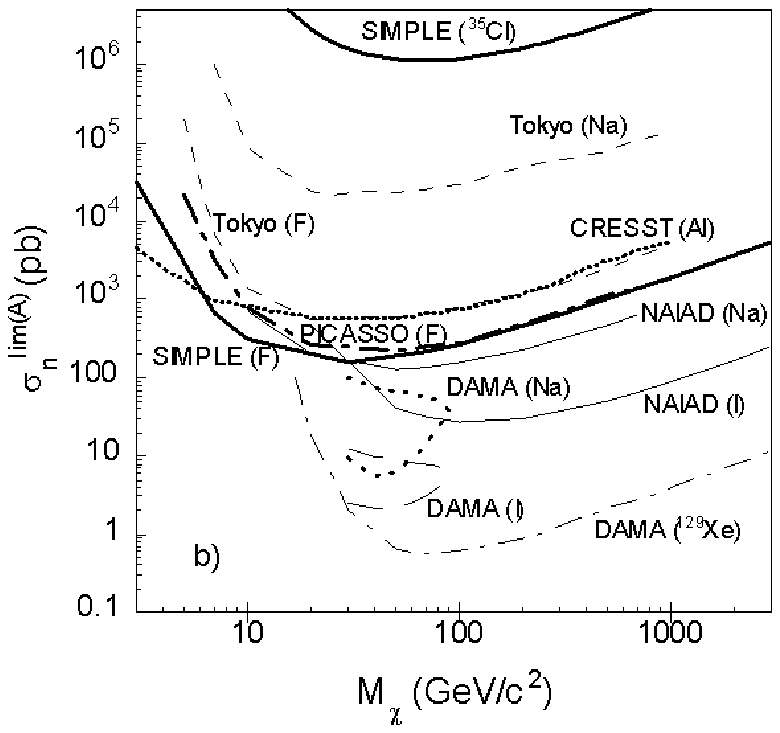}
 \caption{\label{sigps}a) $\sigma_{p}^{lim(A)}$ and b) $\sigma_{n}^{lim(A)}$ vs $M_{\chi}$ for various elements by data from SIMPLE (thick solid), NAIAD (solid), CRESST (dotted), DAMA/Xe-2 (dash-dotted), DAMA/NaI (dotted for Na, dashed for I), and Tokyo (dashed).}
\end{figure*}

\noindent several current experiments (Table 1) in Fig. \ref{sigps}, using spin values from Table 2. The calculation of Na and I $\sigma_{p,n}^{lim(A)}$ for DAMA/NaI require recalculation of $\sigma_{A}^{lim}$ for Na and I from a mixed model \cite{danai} $\sigma_{p}$ and $\sigma_{n}$, followed by re-application of Eqns. (\ref{sigpnlim}). While not completely rigorous, this is a good approximation. The DAMA/Xe-2 experiment, instead, uses almost pure $^{129}Xe$, so it suffices to multiply the $\sigma_{n}^{lim(A)}$ reported in Ref. \cite{damaXe}
by$(\frac{\mu_{p}<S_{n}>}{\mu_{n}<S_{p}>})^2$ to obtain also $\sigma_{p}^{lim(A)}$. CRESST did not report a $\sigma_{n}^{lim(A)}$, but since $^{16}$O is an even-even, spinless and doubly magic nucleus with no magnetic moment \cite{isotable}, its spin-dependent response has been neglected, and the exclusion of Ref. \cite{CRESST} has been assumed a result of Al only.

 \begin{table*}
 \caption{\label{expos}current experiment exposures (detector active mass times measurement duration).}
 \begin{ruledtabular}
 \begin{tabular}{|c|c|c|c|c|c|c|c|}
experiment & NAIAD \cite{naiad} & DAMA/Xe-2 \cite{damaXe} & DAMA/NaI \cite{danai} & PICASSO \cite{picasso} & SIMPLE \cite{plb} & Tokyo/NaF \cite{Tokyo} & CRESST \cite{CRESST}\\
exposure (kgdy) & 3879 & 1763.2 & 57986 & 0.056 & 0.19 & 3.38 & 1.51
 \end{tabular}
 \end{ruledtabular}
 \end{table*}

\begin{table}
\caption{\label{spins} Spin values for relevant nuclides.}
\begin{tabular}{|c|c|c|c|c|c|}
\hline
Nucleus & Z & J & $<S_{p}>$ & $<S_{n}>$ & Ref.\\
\hline
$^{19}$F & 9 & 1/2 & 0.441 & -0.109 & \cite{Tovey}\\ \hline
$^{23}$Na & 11 & 3/2 & 0.248 & 0.020 & \cite{Ressell}\\ \hline
$^{27}$Al & 13 & 5/2 & 0.343 & 0.030 & \cite{ERTO}\\ \hline
$^{35}$Cl & 17 & 3/2 & -0.059 & 0.011 & \cite{Tovey}\\ \hline
$^{37}$Cl & 17 & 3/2 & -0.178 & 0 \footnote{ calculated in the odd group approximation \cite{plb}, using data from Ref. \cite{isotable}. Since $^{37}$Cl and $^{39}$K have the same number of neutrons and similar number of protons, $^{37}$Cl spin values can be evaluated more accurately than in Ref. \cite{plb} by assuming the same neutron group spin and angular momentum as $^{39}$K.} & \cite{plb}\\ \hline
$^{127}$I & 53 & 5/2 & 0.309 & 0.075 & \cite{Ressell}\\ \hline
$^{129}$Xe & 54 & 1/2 & 0.028 & 0.359 & \cite{Ressell}\\ \hline
\end{tabular}
\end{table}

Experiments based on natural Ge and Si are not included, since their only spin-dependent sensitive isotopes are $^{73}$Ge (7.8 \% natural Ge) and $^{29}$Si (4.67 \% natural Si), respectively. Such small isotopic abundances lower the spin-dependent effective mass of detectors based on natural Ge and Si by a factor 10.

\begin{figure}
 \includegraphics[width=8 cm]{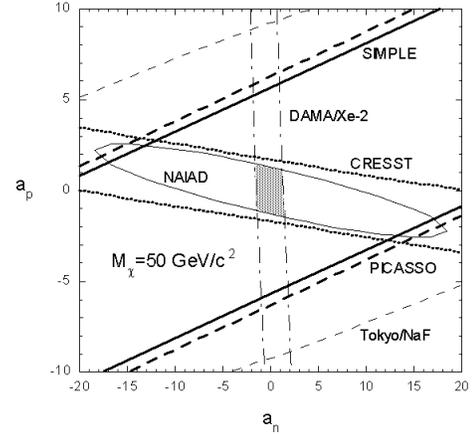}
 \caption{\label{stat}Excluded regions in the $a_{p}-a_{n}$ plane for $M_{\chi} = 50 \text{ GeV/c}^{2}$ , for the experiments of Table \ref{expos}. The resulting allowed region is determined by the intersection of NAIAD and DAMA/Xe-2.}
 \end{figure}

The allowed $a_{p,n}$ for the null result experiments of Fig. \ref{sigps} are shown in Fig. \ref{stat} for $M_{\chi} = 50 \text{ GeV/c}^{2}$. This mass lies in the range 20 to 100 GeV/c$^{2}$, where most experiments reach maximum sensitivity (smallest $\sigma_{p,n}^{lim(A)}$).

\noindent Outside this range, the area of an experimentally allowed ellipse increases. Owing to the difference in sign between the ratios $\frac{<S_{n}>}{<S_{p}>}$ of F and I, the angle between the ellipses of F-based and iodine-based experiments is large, allowing for an intersection significantly smaller than the original ellipses of each experiment. Although DAMA/Xe-2 only provides two parallel lines, these are almost vertical due to the low value of $<S_{p}>$, and the intersection of the DAMA/Xe-2 and the NAIAD ellipses is a very thin slice. Note that CRESST at 1.51 kgdy restricts $a_{p}$ at almost the same level as NAIAD, which has a much larger exposure (3879 kgdy). SIMPLE and PICASSO already exclude a small fraction of the NAIAD allowed ellipse, even with prototype low exposure results. Current limits from the intersection of all these experiments (shaded area) at $M_{\chi} = 50 \text{ GeV/c}^{2}$ are $|a_{n}| \leq 1.3$ and $|a_{p}| \leq 1.5$, and are effectively determined by only NAIAD and DAMA/Xe-2. In terms of cross sections, these are $\sigma_{n} \leq 0.6$ and $\sigma_{p} \leq 0.8$ pb. Consequently, spin-dependent WIMP candidates interacting mainly with protons are excluded at about the same level of those interacting mainly with neutrons. As per the interaction strength, the limits on $a_{p,n}$ from null result experiments do not yet exclude a WIMP candidate with an ordinary weak interaction strength ($|a_{p,n}|=1$).

If an experiment is equally sensitive, at a given $M_{\chi}$, to all possible candidates, its ellipse becomes a circle. This can be seen by rewriting Eqn. (\ref{multi}) as:

\begin{equation}
\alpha a_{p}^{2}+2\beta a_{p}a_{n}+\gamma a_{n}^{2} \leq \frac{\pi}{24G_{F}^{2}\mu_{p}^{2}}
\label{ellissi}
\end{equation}

\noindent where the coefficients $\alpha$, $\beta$, and $\gamma$ are given by:

\begin{equation}
\left\{\begin{array}{l}
\alpha=\sum_{A}{\frac{1}{\sigma_{p}^{lim(A)}}}\\
\beta=\sum_{A}{\pm\frac{1}{\sqrt{\sigma_{p}^{lim(A)}\sigma_{n}^{lim(A)}}}}\\
\gamma=\sum_{A}{\frac{1}{\sigma_{n}^{lim(A)}}}
\end{array}\right .
\end{equation}

\noindent The sign of each term in the $\beta$-summation is given by the sign of the ratio $\frac{<S_{n}>}{<S_{p}>}$ for the isotope of mass number A. The coefficients $\alpha$, $\beta$, and $\gamma$ determine the shape of the ellipse in the $a_{p}-a_{n}$ plane, hence its semiaxes and the angle between the major semiaxis and the $a_{p}$ axis. To make $\beta= 0$ in Eqn. (\ref{ellissi}), corresponding to ellipse axes parallel to the coordinate axes, the detector must contain nuclei with different signs of $\frac{<S_{n}>}{<S_{p}>}$; to get $\alpha=\gamma$ both odd Z and odd N nuclei are needed.

Until a positive signal is found, experimental improvements simply shrink the various ellipses.
In the case of a positive signal, like DAMA/NaI, both an upper ($\sigma_{A}^{lim}$) and a lower ($\sigma_{A}^{lim inf}$) limit on $\sigma_{A}$ exist:

\begin{equation}
\frac{\sigma_{A}}{\sigma_{A}^{lim}} \leq f_{A} \leq \frac{\sigma_{A}}{\sigma_{A}^{lim inf}}
\end{equation}

\noindent where the isotopic fractions $f_{A}$ are again due to the overestimate of attributing the entire counting rate to isotopic specie A only, which affects both $\sigma_{A}^{lim}$ and $\sigma_{A}^{lim inf}$ by a factor$f_{A}^{-1}$. Analogy with Eqn. (\ref{sigpnlim}) then leads to the auxiliary cross sections:

\begin{equation}
\sigma_{p,n}^{lim inf(A)}=\frac{3}{4}\frac{J}{J+1}\frac{\mu_{p,n}^{2}}{\mu_{A}^{2}}\frac{\sigma_{A}^{lim inf}}{<S_{p,n}>^{2}}
\label{sigpnliminf}
\end{equation}

\noindent allowing to complete Eqn. (\ref{multi}) with:

\begin{equation}
\sum_{A}{(\frac{a_{p}}{\sqrt{\sigma_{p}^{lim inf(A)}}} \pm \frac{a_{n}}{\sqrt{\sigma_{n}^{lim inf(A)}}})^{2}} \leq \frac{\pi}{24G_{F}^{2}\mu_{p}^{2}}.
\label{multinf}
\end{equation}

\noindent These extra equations transform the $a_{p}-a_{n}$ plane ellipses into elliptical "shells" and pairs of bands symmetric with respect to the origin. WIMP masses are excluded when the lower limits of Eqn. (\ref{multinf}) are incompatible with the upper limits of Eqns. (\ref{mono})-(\ref{ellissi}).

\begin{figure}
 \includegraphics[width=8 cm]{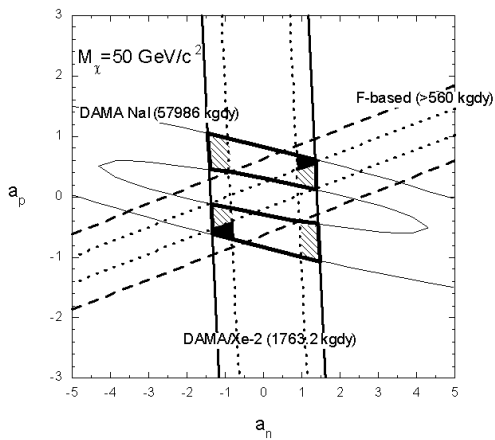}
 \caption{\label{wish}Permitted regions for a positive signal (elliptical shell), and two null signal (dashed and solid) experiments. The dotted inner lines are arbitrary and for illustrative purpose only. The 2 thick border areas are the intersection of current DAMA/NaI and DAMA/Xe-2 results; if also DAMA/Xe-2 had a positive signal this intersection would become the 4 hatched regions; a positive signal F-based experiment could reduce such intersection to the two solid regions (see text).}
 \end{figure}

Fig. \ref{wish} shows the intersection (thick border) of DAMA/NaI and DAMA/Xe-2 permitted regions for $M_{\chi}=50 \text{ GeV/c}^{2}$. NAIAD is reaching this region but to date cannot cut any part of it. When the positive signal of DAMA/NaI is considered, the limits for $M_{\chi}=50 \text{ GeV/c}^{2}$ become $0 \leq |a_{n}| \leq 1.3$, $0.5 \leq |a_{p}| \leq 1.1$, or $0 \leq \sigma_{n} \leq 0.6$, $0.07 \leq \sigma_{p} \leq 0.4$ pb. Though tighter than the above, these limits on the WIMP-nucleon coupling strengths are still compatible with the ordinary weak interaction strength. 

The two allowed areas are symmetric with respect to the origin and correspond to one in the cross section representation. This symmetry and doubling are due to the fact that $a_{p,n} \propto \pm \sqrt{\sigma_{p,n}}$, so that each point in the cross section representation corresponds to two symmetric points in the coupling strength representation.

The inability to set a lower limit for $|a_{n}|$ is due to the null signal of the DAMA/Xe-2 experiment.
Had DAMA/Xe-2 a positive signal, its permitted region would become a pair of bands symmetric with respect to the origin of the $a_{p}-a_{n}$ plane, leading to the four intersection regions (two in the cross section representation) shown in Fig. \ref{wish} as hatched. To reduce the number of allowed regions, a third experiment with a different orientation is needed, such as one of the F-based experiments of Fig. \ref{stat}. Fig. \ref{wish} also contains a projection for PICASSO (dashed), along with an hypothetical positive signal (dotted); other F-based experiments require improvements or higher exposures to reach the same level. Note that F-based experiments seem competitive even with exposures significantly lower than sodium iodide-based.

In the case of Fig. \ref{wish}, the F-based experiments would remove completely two of the hatched areas, allowing only the two solid regions symmetric with respect to the origin. The two regions correspond to positive (repulsive) and to negative (attractive) WIMP-nucleon interaction.

If there is a single WIMP specie, there must be a pair of symmetric regions (one in the cross section representation) permitted by all experiments with positive signal. If there is no such common permitted region, the existence of a single WIMP specie is excluded.


%



\begin{acknowledgments}
I wish to thank my colleagues of the SIMPLE collaboration for their encouragement and support, and the PICASSO collaboration for its hospitality during my recent visit. This work has been supported by grant POCTI/FNU/39067/2002 of the Portuguese Foundation for Science and Technology (FCT), co-financed by FEDER. The author is supported by grant SFRH/BPD/13995/2003 of FCT.
\end{acknowledgments}


\end{document}